\documentclass[fleqn,twoside]{article}
\usepackage[headings]{espcrc2}

\readRCS
$Id: espcrc2.tex,v 1.2 2004/02/24 11:22:11 spepping Exp $
\usepackage{graphicx}
\usepackage{epstopdf}

\newcommand {\be} {\begin{equation}}
\newcommand {\ba} {\begin{eqnarray}}
\newcommand {\ee} {\end{equation}}
\newcommand {\ea} {\end{eqnarray}}

\usepackage[figuresright]{rotating}

\newcommand{\AmS}{{\protect\the\textfont2
  A\kern-.1667em\lower.5ex\hbox{M}\kern-.125emS}}

\title{Hadron Form Factors in AdS/QCD\thanks{Invited talk at the Eighth International Conference on Hyperons, Charm and Beauty Hadrons (BEACH 2008), Columbia, SC, USA, 22--28 June 2008.}
}

\author{Carl E.\ Carlson\address[MCSD]{Physics Department,
        College of William and Mary,  \\ 
        P.O. Box 8795, Williamsburg, VA 23187-8795, USA}%
         }

\runtitle{Hadron Form Factors in AdS/QCD}
\runauthor{C.\ E.\ Carlson}

\begin{document}

\begin{abstract}
We discuss how to calculate form factors using a holographic model of
QCD,  mainly focusing on vector, axial, and pseudoscalar mesons.   We
illustrate the techniques on gravitational form factors (which are
useful for constraining the generalized parton distributions) as well as
quoting the results for electromagnetic form factors.  One striking
outcome, comparing the several types of calculated form factors, is that
mesons appear noticeably more compact when measured by the gravitational
form factors than when measured by the charge form factor.
\vspace{1pc}
\end{abstract}

\maketitle

\section{INTRODUCTION}

Jim Gates, writing in \textit{Physics Today}~\cite{Gates:2006bv} described the AdS/CFT correspondence as ``an unexpected link between gauge theories like QCD and gravitational theories'' that ``may be used to carry out high-precision calculations in gauge theories.''   Since gauge theories underlie all the physics we are interested in, this is a powerful claim, and one that we should understand, use (if possible), and assess.

It will be the goal of this talk and this write-up to explain the vocabulary of ``AdS'' and ``CFT'', to explain how calculations in one of those arenas can be used to obtain results in the other, and to apply the ideas of the AdS/CFT correspondence, or its morph into an AdS/QCD correspondence, to obtain results for hadronic form factors.   The description of results will concentrate on meson form factors, both electromagnetic~\cite{Grigoryan:2007vg,deTeramond:2005su} and gravitational~\cite{Abidin:2008ku,Brodsky:2008pf}.  The latter are in fact experimentally accessible, not because anyone dreams of doing gravitational scattering with elementary particles, but because of their direct connection to generalized parton distributions.   Results have also been obtained for nucleon form factors using AdS/CFT, which we will simply cite~\cite{Hong:2007kx}.  To the best of my knowledge, there have been no AdS/CFT applications to heavy quark transition form factors, but perhaps this conference will stimulate work in that direction.

The idea of an AdS/CFT correspondence started with Maldacena's conjecture~\cite{Maldacena:1997re} of a correspondence between a certain type of string theory (type IIB, specifically) that existed in 10 dimensions and the four dimensional version of supersymmetric pure Yang-Mills theory.  The latter is the ``conformal field theory'' (CFT) of the acronym.  The Yang-Mills theory has two parameters.  One is $N_C$, the number of colors that take part in the interactions, and the other is $g_{Y\!M}$, the coupling strength of the Yang-Mills theory, which is also the coupling to matter in a theory where quarks are introduced.  That there is a correspondence means that certain quantities such as propagators or matrix elements which were calculated in one theory would be just the same in the other.  The correspondence can be most useful if, as may actually be the case, quantities that can be calculated in weak coupling in one theory will be valid for strong coupling in the other.

Working with string theory to obtain quantities relevant phenomenologically is not currently feasible.  So there begins a series of changes and approximations~\cite{Erdmenger:2007cm} that will help connect a weakly coupled multidimensional field theory to QCD, the theory of real world strong interactions.  Consider the ``low energy'' limit of string theory.  The string appears pointlike and has particle-like excitations;  the still supersymmetric theory that ensues is supergravity.  There is more detail, in that the string approaches its classical limit when the product 
$N_C g^2_{Y\!M}$ gets very large.  Still further, in the low energy limit the 10 dimensional space can be split into a product of and 5-dimensional (5D) AdS or AdS$_5$ space and a 5D sphere $S^5$.   We give the sphere a small radius, so that it may be eliminated from current phenomenological consideration by saying that the $S^5$ part of any wave function is in the ground states, since excitations would be very massive.   We are left with a 5D gravitational theory on AdS$_5$ (still to be defined).

On the CFT (also still to be defined) side, we have a theory that is not QCD.  However, it has been argued that QCD is approximately conformal over certain kinematic 
ranges~\cite{deTeramond:2005su}, and that there will be a connection between some gravitational theory in higher dimensions that includes matter and QCD.

The foregoing has been stated by way of motivation.  There is not a precise \textit{ab initio} correlation currently known between a string theory and QCD, but one thinks that one may exist. It was the idea of Erlich \textit{et al.}~\cite{Erlich:2005qh} and of Da Rold and Pomarol~\cite{Da Rold:2005zs} to start in the middle, and ask, after one had reduced the 10D theory down to a gravitational theory on AdS$_5$, what terms could possibly exist in the Lagrangian.  If one states what strongly-interacting particles one is interested in and limits oneself to the simplest terms, there are not many possibilities.  One can then study the phenomenological consequences for the 4D correspondent theory, and see if the results accord with our observed universe.  This way of proceeding is often referred to as a ``bottom-up'' approach to the AdS/CFT correspondence, or as AdS/QCD.

To proceed, we will explain a few things more mathematically, and then discuss form factors.

\section{DEFINITIONS}

Five dimensional Anti-de Sitter space, AdS$_5$, is (to me) most easily thought of as a five dimensional hyperboloidal surface embedded in a 6D space, and given by
\be
t^2 - y_1^2 - y_2^2 - y_3^2 - y_5^2 + y_6^2 = L^2 (= 1)  .
\ee
The embedding space has a metric given by
\be
ds^2 = dt^2 - dy_1^2 - dy_2^2 - dy_3^2 - dy_5^2 + dy_6^2.
\ee
One notices that the surface has an $SO(4,2)$ symmetry, in the same sense that Lorentz invariance is an $SO(3,1)$ symmetry.  The AdS surface has constant negative curvature, and one can view the space as having four ordinary Minkowski coordinates, plus two extra.  Commonly, so that we have only a five dimensional space to deal with, we change variables, so that the ordinary coordinates are $x^\mu$ and the extra coordinate is (risking confusion) $z$, with $0<z<\infty$,
\ba
ds^2 &=& \frac{L^2}{z^2} 
\left( dx_\mu dx^\mu - dz^2 \right)  \nonumber \\
&=& \frac{L^2}{z^2} 
\left( dt^2 - d\vec x^2 - dz^2 \right)  \,.
\ea

To obtain the conformal group, one starts with the usual translations and Lorentz transformations, or the Poincar\'e group,
\ba
x_\mu &\to& x'_\mu = x_\mu + a_\mu  \,,
\nonumber  \\
x_\mu &\to& x'_\mu 
    = \Lambda_\mu^{\ \nu} x_\nu   \,,
\ea
and then includes the dilation, or simple expansion, operation
\be
x_\mu \to x'_\mu = \lambda\  x_\mu  \,,
\ee
where $\lambda$ is an ordinary real number.  By commuting the generators of the Poincar\'e group and of dilations one gets four new generators, for a total of 15,  which generate what are called special conformal transformations.   These 15 generator form a group called the ``conformal group,''  and the group has the same algebra as $SO(4,2)$.   A field theory which is invariant under the conformal group is a conformal field theory.

There is an immediate problem in connecting to the real world, in that if one can dilate the coordinates arbitrarily, then one can dilate the momentum arbitrarily, and thus the momentum-squared or mass-squared.  Hence a conformal field theory has either a continuous mass spectra, or has only massless particles.  Additionally, there should not be any mass scale associated with the coupling parameter, \textit{i.\,e.,} it should not run.  This is not QCD.

However, one can make the AdS theory also a little non-$SO(4,2)$ invariant by, for the simplest example, putting an upper limit $z_0$ upon the range of $z$.  This will have the effect of quantizing the masses of the particles one finds in the 4D theory.  One can also argue that QCD is not so far from conformal, at least within a limited kinematic range.  For example, the coupling parameter measured from the spin structure function at low momentum transfer does not seem to run quickly~\cite{Deur:2008rf}.

Thus, even if the original AdS/CFT correspondence idea is correct, there are a number of approximations along the way to applying it to get results for QCD, and the seriousness of the approximations is not quantified.  However, we would like to see what the results look like, testing them at first and then perhaps pressing into areas where there are not other means to obtain strong coupling QCD results.  First we need to know more about the details of the correspondence.

\section{THE AdS/QCD CORRESPONDENCE}

To give the correspondence, we need to review how one can calculate the vacuum expectation values of operators from a generating function.   In the 4D space, let $Z_{4D}$ be the generating function, let ${\mathcal O}(x)$ be a typical operator involved in the expectation value, and let $\phi^0(x)$ be the source function that goes along with $\mathcal O$.  Then
\ba
\left\langle {\mathcal O}(x) \dots
\right\rangle  =  
-i \frac {\delta Z_{4D}}
{\delta \phi^0(x) \ldots }	\,,
\ea
with
\ba
Z_{4D}[\phi^0] = \left\langle \exp
	\Big( iS_{4D} + i \int d^4x \ {\mathcal O(x)} \phi^0(x) \Big) \right\rangle	\,,
		\nonumber
\ea
where $S_{4D}$ is the action for the 4D theory.

The operational rule for the AdS/CFT correspondence is that the action for the 4D theory is the same as the classical action for the 5D theory,
\be
Z_{4D}[\phi^0] = \exp \big( iS_{5D}[\phi_{cl}] \big)	\,,
\ee
where $\phi_{cl}(x,z)$ is a solution to the 5D equations of motion with boundary condition
\be
\lim_{z \to 0} \phi_{cl}(x,z) = \phi^0(x)	\,.
\ee


\section{APPLICATION TO FORM FACTORS}


Electromagnetic form factors are defined in terms of matrix elements of the electromagnetic current.  Using spin-1 rho mesons for starters, the matrix elements we want to calculate are
\be
\left<\rho_n^a(p_2,\lambda_2)\big|J^{\mu}(0)\big|\rho_n^b(p_1,\lambda_1)\right>  \,,
\ee
where $J^\mu(x)$ is the electromagnetic current and $\rho_n^b(p_1,\lambda_1)$ is the n$^{th}$ recurrence of a rho meson with isospin $b$ and the stated momentum and helicity.  Similarly, gravitational form factors are defined in terms of matrix elements of the stress or energy-momentum tensor $\hat{T}^{\mu\nu}(x)$,
\be
\label{eq:stressmatrix}
\left<\rho_n^a(p_2,\lambda_2)\big|\hat{T}^{\mu\nu}(0)\big|\rho_n^b(p_1,\lambda_1)\right>
\ee

Generally speaking, taking the stress tensor case for definiteness, one calculates the three point function, or vacuum expectation value
\ba
\big< 0 \big|{\cal T} \big( {J_a}^\alpha(x)  T^{\mu\nu}(y)  {J_b}^\beta(w)  \big)\big|0\big>  \,,
\ea
using the current farthest to the right to create the desired hadronic state (and with a little care, obtaining only the desired state), the current farthest to the left takes the final hadron state back to the vacuum, so one obtains just the desired matrix element of the stress tensor,~(\ref{eq:stressmatrix}).   

In order to get the correct normalizations for the states, we first should look at the two-point functions or propagators.


\subsection{Two-point Functions}


It is time to give some details of the 5D Lagrangain or action we shall use, as well as some details of the connections between 4D operators and the 5D sources or fields they correspond to.  The objects we will focus on in 4D space are vector currents (although the axial vector current poses no problem here) and the stress tensor.  The correspondences are,
\ba
\Big({\mathcal O}(x) &\leftrightarrow& \phi(x,z)   \Big) \,, \\
\bar q \gamma^\mu t^a q = 
{J^a}^\mu(x) &\leftrightarrow& {V^a}^\mu(x,z)  \,,  \\
T_{\mu\nu}(x)  &\leftrightarrow& h_{\mu\nu}(x,z)  \,.
\ea
The vector current is ${J^a}^\mu(x)$.   Its quark representation is listed to show what it would be with explicit quarksl  $t^a$ is an isospin matrix.  The vector field is  ${V^a}^\mu(x,z)$, with $a$ an isospin index, and $h_{\mu\nu}(x,z)$ is the fluctuation in the metric,
\be
g_{\mu\nu}(x,z) = \frac{1}{z^2} \left( \eta_{\mu\nu} + h_{\mu\nu}(x,z) \right)  \,,
\ee
where $\eta_{\mu\nu} = diag(1,-1,-1,-1,-1)$ is the flat space metric.

Still focusing on the vector fields and gravity (which gives the AdS space),  the 5D action is
\ba
S_{5D}=\int d^5 x \sqrt{g}\bigg\{ \mathcal{R}+12	-\frac{1}{4g_5^2}
V^{MNa} V^a_{MN}  \bigg\}
\ea
where $g$ is the determinant of the metric tensor, $\mathcal R$ is the curvature scalar, $g_5$ is a coupling parameter, and $V^a_{MN}$ is the field
\be
V_{MN}=\partial_M V_N -\partial_N V_M-i[V_M,V_N]
\ee
(although the Yang-Mills part is not needed for $n$-point functions we calcualate).  The ``$12$'' is the cosmological constant which gives the constant curvature AdS$_5$ space.

To get the classical solutions for the 5D fields, we solve the equation of motion for $V_\mu$, which comes from the Euler-Langrange equation in the usual way, and in momentum space for the 4D coordinates reads
\be
\label{eq:eom}
\left(   
z \, \partial_z \left( \frac{1}{z} 
\partial_z V^a_\mu(q,z) \right)
+  \, q^2 
V^a_\mu(q,z)   \right)_\perp
 = 0
\ee 
in a gauge where $V_5 = V_z = 0$ and $\partial^\mu V_\mu = 0$.  The equation is the same for all values of the index $\mu$, so that one can factor the solutions as
\be
{V_\perp}_\mu(q,z)=V(q,z) V^0_\mu(q)
\ee
where $V(q,z)$ is the ``profile function'' or ``bulk-to-boundary propagator,'' and satisfies the same equation as $V_\mu(q,z)$.  The $z=0$ boundary condition must be $V(q,\epsilon) = 1$ (the limit $\epsilon\to 0$ is implied) and we impose a Neumann boundary condition, $\partial_z V(q,z_0) = 0$, at the other end.  The equation allows an analytic solution in terms of Bessel functions $J_1$ and $Y_1$,  and the profile function is
\ba
V(q,z)=\frac{\pi}{2}zq\left(\frac{Y_0(qz_0)}{J_0(qz_0)}J_1(qz)-Y_1(qz)\right)		\,.
\ea

A useful alternative is to expand the profile function, as is allowed by the Sturm-Liouville theorem,  in terms of normalizable solutions $\psi_n(z)$ to an equation like Eq.~(\ref{eq:eom}),
\be 
z \, \partial_z \left( \frac{1}{z}  \partial_z \psi_n(z) \right)
+  \, m_n^2  \psi_n(z)     = 0    \,.
\ee
The boundary conditions are $\psi_n(0) = 0$ and $\partial_z \psi_n(z_0) = 0$ and the  normalization condition is $\int_0^{z_0}  (dz/z)  \psi_m(z) \psi_n(0) = \delta_{mn}$.  The solutions are $\psi_n(z) = const. \times z J_1(m_n z)$  and the high $z$ boundary condition fixes the discrete eigenvalues $m_n$ from $J_0(m_n z_0) = 0$.   With some manipulation, one can show that
\be
V(q,z)= - g_5\sum_n \frac{F_n\psi_n(z)}{q^2-m_n^2}  \,,
\ee
where $F_n$ is a constant defined from
\be
\lim_{z\to 0} \frac{1}{z} \partial_z \psi_n(z) = g_5 F_n \,.
\ee

Inserting the solution for $V^a_M$ back into the action leads to a result that comes only from a surface term at $z=0$.  The part quadratic in $V$ is
\ba
S_{5D}  \stackrel{\to}{=}  \int \frac{d^4 q}{(2\pi)^4} 
   {V^0}^\mu(q){V^0}_\mu(q)\left(-\frac{\partial_z V(q,z)}{2g_5^2z} \right)_{z=\epsilon} .
   	\nonumber
\ea
The differentiation needed, by the AdS/CFT correspondence, to get the current-current 2-point function is
\be
\label{eq:2pt}
\left<0 \right| {\mathcal T} 
J^{a\mu}(x)J^{b\nu}(y) 
\left| 0 \right>
= -i
\frac{  \delta^2 S_{5D}}
{\delta V^{a0}_\mu (x)  \delta V^{b0}_\nu(y)}	,
\ee
and this by and by leads to
\ba
 i \int d^4 x \, e^{iqx} \left< 0 \right| {\mathcal T} J_\mu^a(x)J_\nu^b(0) \left| 0 \right>  &&
	\nonumber \\
 	=\sum_n \frac{F_n^2 \, \delta^{ab}}{q^2-m_n^2+i\varepsilon}
		\left(\eta_{\mu\nu}  -  \frac{q_\mu q_\nu}{q^2}\right)   .    \hskip - 5 em && 
\ea
($\mathcal T$ is the time-ordering operator.)  This is by itself a remarkable result:  the poles in the 4D variable $q^2$ have been determined by the eigenvalues of a 5D equation.

Numerically, using the experimental mass of the rho as the mass of the lightest spin-1 state leads to $1/z_0 \approx 0.32 {\rm\ GeV}$.

We close this section by noting that an alternative definition of $F_n$ can be given using the matrix element of the current,
 \be
 <0|J_\mu^a(0)|\rho_n^b(p)>=F_n\delta^{ab}\varepsilon_\mu(p)  \,,
 \ee
where $\varepsilon_\mu(p)$ is a polarization vector.  Using this definition, and evaluating the 2-point function in Eq.~(\ref{eq:2pt}) by inserting a complete set of states, shows that this definition agrees with the previous one.


\subsection{Three-point Functions}


For the three-point function, needed to isolate the matrix element that gives the form factor, we take the derivatives indicated in
\ba
\label{eq:3pt}
\big< 0 \big| {\mathcal T} {J}^\alpha(x) \hat{T}^{\mu\nu}(y){J}^\beta(w)\big| 0 \big>	&&
					\nonumber \\[1.2ex]
= \frac{  -  2 \, \delta^3 S_{5D}}
	{\delta V^0_\alpha (x) \delta h^0_{\mu\nu}(y) \delta V^0_\beta(w)}  \,.  \hskip - 6 em &&
\ea
In the action we need just the terms that involve the metric tensor fluctuation $h_{\mu\nu}$ once and the vector field twice,  which are contained in 
\be
S_{5D} \stackrel{\to}{=} - \frac{1}{4g_5^2}\int d^5 x \sqrt{g} \,  g^{lm}g^{pn}V^a_{mn}V^a_{lp}  \,.
\ee
[Recall that $g_{\mu\nu} = (\eta_{\mu\nu} + h_{\mu\nu})/z^2$.]

Before continuing, one will have to analyze the 2-point function for a pair of $h_{\mu\nu}$'s, similarly to what we did for the vector field in the last subsection.  We will only quote the results, which are that, at least for the transverse-traceless part of $h_{\mu\nu}$, the result also factors,
\be
{h}_{\mu\nu}(q,z)=h^0_{\mu\nu}(q)   {\mathcal H}(q,z)   \,,
\ee
and the profile function is
\ba
\mathcal{H}(Q,z)=\frac{1}{2}Q^2z^2
		\bigg(\frac{K_1(Qz_0)}{I_1(Qz_0)}I_2(Qz)+K_2(Qz) \bigg) .    \nonumber
\ea
We quoted the profile function in a form suitable for spacelike $q^2 = - Q^2$, and $I_2$ and $K_2$ are the modified Bessel functions.


\subsection{Form-Factor Results}


From the 3-point function, Eq.~(\ref{eq:3pt}), it is possible to isolate contributions from individual spin-1 states.  We will quote the results in terms of the form factors.  There are 6 form factors for the stress tensor, or 6 gravitational form factors, in the spin-1 case,
\ba
\left<\rho_n^a(p_2,\lambda_2)\big| {T}^{\mu\nu}(0)\big|\rho_n^b(p_1,\lambda_1)\right>
= \varepsilon^*_{2\alpha} \varepsilon_{1\beta}   &&
						\nonumber \\[1ex]
\times	\Big\{ -2 A(q^2) \eta^{\alpha\beta} p^\mu p^\nu  \hskip 7.5 em   &&
						\nonumber \\
-\ 4 \big(  A(q^2)+B(q^2)  \big)  q^{[\alpha} \eta^{\beta](\mu} p^{\nu)}
	+ {\rm \ 4\ more} \Big\}	\hskip - 2 em			&&
\ea
where, as it happens, the form factors $A$ and $B$ depend only on the transverse-traceless part of the stress tensor.  

The results for $A$ and $B$ are
\ba
A(q^2) &=& \int_0^{z_0} \frac{dz}{z} \mathcal{H}(Q,z) \psi_n(z) \psi_n(z)
			 \nonumber  \\[1ex]
B(q^2) &=& 0
\ea

\begin{figure}[tbp]
\vspace { - 6 mm }
\begin{center}
\includegraphics[width=7cm]{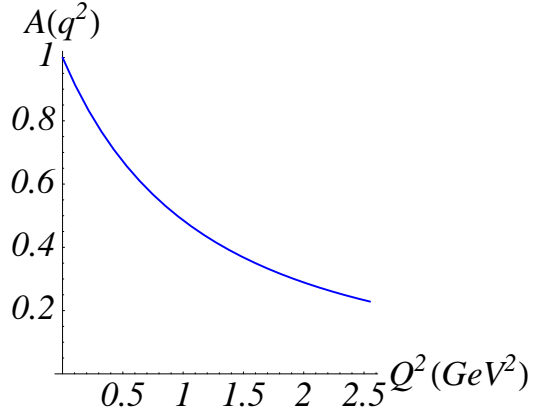}
\end{center}
\vspace{ - 13 mm }
\caption{Gravitational form factor $A(q^2)$ from AdS/QCD, for the lightest spin-1 state.}
\label{fig:one}
\end{figure}

There is one scale in the problem, and we have already fixed it from a physical quantity, \textit{viz.}, $1/z_0 \equiv \Lambda_{\rm QCD} = 0.32$ GeV.  We  plot $A(q^2)$ for the lightest of the spin-1 states in Fig.~\ref{fig:one}.  The radius that obtains for this form factor is~\cite{Abidin:2008ku},
\be
\left\langle r^2 \right\rangle_{\rm grav} = -6 \left. \frac{\partial A}{\partial Q^2} \right|_{Q^2=0}
= \frac{3.24}{m_\rho^2} = 0.21 {\rm\ fm}^2   .
\ee
In a similar fashion, one can use AdS/QCD to obtain the electromagnetic form factors of spin-1 particles~\cite{Grigoryan:2007vg}.  The charge radius works out as
\be
\left\langle r^2 \right\rangle_{\rm C} = -6 \left. \frac{\partial G_C}{\partial Q^2} \right|_{Q^2=0}
= 0.53 {\rm\ fm}^2   \,.
\ee

That is, the size of the particle measured by the gravitational form factor is noticeably less than the charge radius.   The form factor $A(q^2)$, Fourier transformed into coordinate space, measures the distribution of the longitudinal component (in light front variables) of the particle's momentum~\cite{Abidin:2008sb}.  Hence the result, more specifically, is that the momentum density of the particle has more compact distribution than the charge density.


\section{FINAL COMMENTS}


We have seen how form factors can be calculated using a connection between 5D theories with gravitational interactions and strongly coupled 4D conformal or QCD-like theories, and using the case of spin-1 rho-meson-like states as a starting situation to stude.  The extensions to spin-1 axial states and to pseudoscalars is also known~\cite{Grigoryan:2007vg,Kwee:2007dd}   Masses, decay constants, and charge radii that can be compared to experimental data are all right at the 10\% or so level.

We described in this talk the less-fancy AdS/QCD approach sometimes called the ``bottom up'' approach.  This involves analyzing what terms must appear in the Lagrangain in the 5D AdS space, and finding the phenomenological implications of this 5D theory when mapped into 4D space using the AdS/CFT correspondence.   One will utimately also want to obtain these terms, if possible, starting from a string theory and taking a low energy limit.  Work in these directions is also under way~\cite{Sakai:2004cn}.

We showed how the calculation of two-point functions---propagators with interactions---proceeds to obtain masses and decay constants.  We obtain form factors from three-point (vertex) functions  Results for electromagnetic and gravitational form factors were briefly presented.  

One interesting result is that articles appear smaller viewed gravitationally than electromagnetically.  The matter within a particle when momentum weighted is more concentrated than when charge weighted.

A potentially interesting future project is to apply the AdS/CFT correspondence to the form factors for flavor changing reactions.  Handling heavy flavors within the AdS/CFT correspondence is, however, not a settled procedure; one work in this direction is in
~\cite{Shock:2006qy}.

\end{document}